\documentclass[twocolumn,fleqn,nobibnotes]{revtex4}

\usepackage{amsmath}
\usepackage{graphicx}
\usepackage{float}
\usepackage{subfigure}

\def\lsim{\raise0.3ex\hbox{$<$\kern-0.75em\raise-1.1ex\hbox{$\sim$}}}
\def\gsim{\raise0.3ex\hbox{$>$\kern-0.75em\raise-1.1ex\hbox{$\sim$}}}

\renewcommand{\vec}[1]{\boldsymbol{#1}}
\newcommand{\dif}{\mathrm{d}}

\begin{document}

\title{Nuclear DVCS at small-$x$ using the color dipole phenomenology}
\author{Magno V.T. Machado}
\affiliation{Centro de Ci\^encias Exatas e Tecnol\'ogicas, Universidade Federal do Pampa \\
Campus de Bag\'e, Rua Carlos Barbosa. CEP 96400-970. Bag\'e, RS, Brazil}

\begin{abstract}
Using the high energy color dipole formalism, we study the coherent and incoherent nuclear DVCS process, $\gamma^* A \rightarrow \gamma\,X$, in the small-$x$ regime. We consider simple models for the elementary dipole-hadron scattering amplitude that captures main features of the dependence on atomic number $A$, on energy and on momentum transfer $t$. Using the obtained amplitudes we make predictions for the nuclear DVCS cross section at photon level in the collider kinematics.

\end{abstract}

\maketitle

\section{Introduction}

An interesting way of probing hadronic matter involves the physics of deeply virtual Compton scattering (DVCS), $ep\rightarrow ep\gamma$, where a parton in the proton absorbs the virtual photon, emits a real photon and the proton ground state is restored. Accordingly, the relevant QCD diagrams for such a process involves the exchange of two gluons at low $x$ (at collider experiments) or two quarks at larger $x$ (as in fixed target experiments) carrying different fractions of the initial proton momentum (skewedness). The DVCS process thus measures generalized parton distributions (GPDs) which depends on two momentum fractions $x$ and $x^{\prime}$, as well as on $Q^2$ and the four-momentum transfer $t$ at the proton vertex. It interferes with the Bethe-Heitler scattering allowing parton scattering amplitudes to be measured. Similar process also occurs in $eA$ colliders, $eA\rightarrow eA \gamma$, which is extremely sensitive to the corresponding nuclear parton distributions. Therefore, DVCS on nuclei will shed light on the understanding of partonic matter in nuclei and on basic questions of QCD regarding the existence of a saturated gluon state, the Color Glass Condensate and the relationship of nuclear Glauber-Gribov shadowing to hard diffraction.

Experimentally, DVCS on nucleons has been studied at high energies by H1 \cite{H1dvcs} and ZEUS \cite{ZEUSdvcs} Collaborations at DESY-HERA and at low energies in the experiment CLAS \cite{CLAS} at the Jefferson Laboratory (JLAb). Initial experimental investigations of nuclear DVCS has been reported by CLAS collaboration also has reported DVCS on deuterium, both at low energies. The interference of DVCS and bremsstrahlung leading to a beam-charge asymmetry has been investigated by HERMES Collaboration \cite{HERMES} at DESY-HERA for a hydroegn and a deuterium target.  At high energies, it is expected to investigate nuclear DVCS on future electron-ion colliders (EICs) and on ultraperipheral $AA$ collisions \cite{YRUPC}. Among the planned $eA$ collides we recall the eRHIC project and the more recent LHeC proposal \cite{LHeC}.
The LHeC is a proposed colliding beam facility at CERN, which will exploit large energy and intensity provided by the LHC for lepton-nucleon (or nucleus) scattering. The existing 7 TeV LHC proton or ion beam will collide with a electron beam simultaneously with proton-proton or heavy ion collisions taking place at the LHC experiments. One purpose is a electron beam circulating in the existing LHC tunnel with a nominal energy of 70 GeV, resulting in a lepton-nucleon scattering with center of mass energy of 1.4 TeV and very luminosity. The large energy and the luminosity would allow the parton densities of the proton (or nucleus) to be measured at unexplored momentum transfers $Q^2$ and small Bjorken $x\leq 10^{-6}$ for
 $Q^2\approx 1$ GeV$^2$. This newly accessed low $x$ regions is ideal to search for QCD dynamics associated to the extremely high partons densities, the so called parton saturation physics.

Nuclear DVCS is the golden exclusive channel for the investigation of the partonic structure of hadrons within the universal framework of GPDs. For the DVCS reaction off nuclei one has two channels: the coherent and the incoherent case. Coherent DVCS corresponds to the channel where the final state consists exclusively of the initial nucleus, which probes directly the nuclear GPDs that may be reconstructed from the elementary nucleon GPDs depending on specific models of the nuclear structure. In the incoherent reaction, we have the break-up of the final nucleus. Thus, for incoherent scattering break-up configurations for the final nucleus into an outgoing nucleon and an $A-1$ system have to  be considered.  A contribution to this break-up process involves a nucleon which is expelelled from the initial nucleus. Such a  process may be described in the impulse approximation as the interaction of the virtual photon with a quark belonging to a nucleon embedded in the nuclear medium. The main theoretical approach to nuclear GPDs for Bjorken $x>0.1$ assumes that they are given by the convolution of unmodified or modified nucleon GPDs with the distribution of nucleons in the nuclear target obtained from the non-relativistic nuclear wave function. For the deuteron case, the GPD formalism was originally developed in Ref. \cite{Cano}, whereas for the case of nuclei having spin 0, 1/2 and 1 it was presented in Ref. \cite{Kirchner}. At the small-$x$ limit, the convolution formalism is not reliable anymore as nuclear shadowing and anti-shadowing effects become important and a model of nuclear GPDs for heavy nuclei in that region was proposed in Refs. \cite{Freund1,Freund2}. In addition, the interplay the coherent and the incoherent contributions to nuclear DVCS was studied in Ref. \cite{Guzey1} and the role of the neutron contribution to nuclear DVCS observables was addressed in \cite{Guzey2}. The medium modifications of the bound proton GPDs and their influence on incoherent DVCS on nuclear targets was computed recently in Ref.. \cite{Guzey3}. In addition, the nuclear GPDs formalism has been compared to framework of generalized vector meson dominance (GVDM) model in Ref. \cite{Guzey4}.

In this work, an alternative theoretical formalism will be considered. We use the high energy color dipole approach \cite{dipole} to study the nuclear DVCS process at photon level. In order to do so, recent phenomenological models for the elementary dipole-hadron scattering amplitude that captures main features of the dependence on atomic number $A$, on energy and on momentum transfer $t$ are considered. This investigation is directly related and complementary to the conventional partonic description of nuclear DVCS, which considers the relevant nuclear GPDs.  The color dipole approach provides a very good description of the data (For a recent review, see \cite{fss}) on $\gamma p$ inclusive production, $\gamma \gamma$ processes, diffractive deep inelastic and vector meson production. In particular, the cross section for DVCS on nucleons is nicely reproduced in several implementations of the dipole cross section at low $x$ \cite{dvcs1,FM,KMW,MPS,Watt,MW}. In particular, the dipole models considering parton saturation effects are very interesting as experimentally the photon virtuality is not always very large and large (at small-$x$) higher-twist corrections can take place and perturbative collinear factorization is expected to break down. For instance, these issues has been addressed in  Ref. \cite{Guzey5} within the framework of the Color Glass Condensate (CGC), where the quark and gluon GPDs at small-$x$ and for nuclear targets are evaluated.

This paper is organized as follows.  In next section, we make a short summary on the color dipole approach applied to the DVCS on nucleons and on nuclei at high energies. In particular, the coherent and incoherent contributions to nuclear DVCS process are investigated in detail.  In Sec.  3, we present the main results for such a process using nuclear targets relevant for the planned EICs.  In last section we summarize the results.

\section{DVCS process on nucleons and nuclei in the color dipole framework}

\begin{figure}[t]
\includegraphics[scale=0.47]{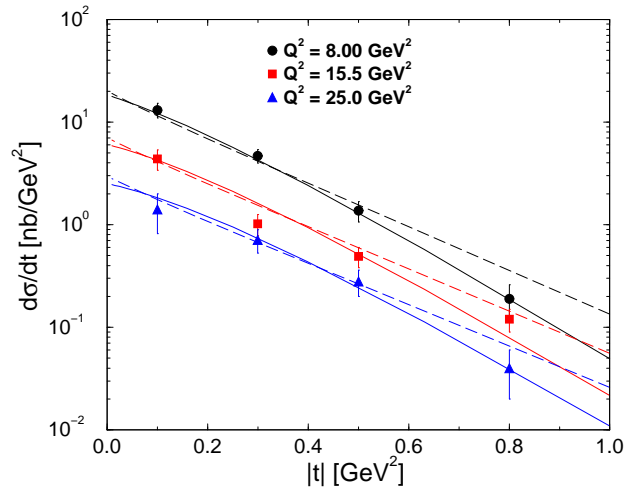}
\caption{The DVCS differential cross section, $d\sigma^{\gamma^*p\rightarrow \gamma p}/d|t|$, as a function of momentum transfer $|t|$ compared to data from DESY-HERA at $W=82$ GeV.}
\label{fig:1}
\end{figure}

In what follows, we summarize the relevant formulas in the color dipole picture for the DVCS process on nucleons and nuclei. In the dipole framework \cite{dipole}, the scattering process $\gamma^* p\rightarrow \gamma p$ is assumed to proceed in three stages: first the incoming virtual photon fluctuates into a quark--antiquark pair, then the $q\bar{q}$ pair scatters elastically on the proton, and finally the $q\bar{q}$ pair recombines to form a real photon. We will analyze the nuclear case later on. The imaginary part of the scattering amplitude for DVCS on nucleons is given by \cite{MPS,KMW,MW}
\begin{eqnarray}
 \mathcal{A}^{\gamma^* p\rightarrow \gamma p}(x,Q,\Delta) & = & \sum_f \sum_{h,\bar h} \int\!\dif^2\vec{r}\,\int_0^1\!\dif{z}\,\Psi^*_{h\bar h}(r,z,0)\nonumber \\
& \times & \mathcal{A}_{q\bar q}(x,r,\Delta)\,\Psi_{h\bar h}(r,z,Q)\,,
  \label{eq:ampvecm}
\end{eqnarray}
where $\Psi_{h\bar h}(r,z,Q)$ denotes the amplitude for a photon (with virtuality $Q$) to fluctuate into a quark--antiquark dipole with helicities $h$ and $\bar h$ and flavor $f$. The quantity $\mathcal{A}_{q\bar q}(x,r,\Delta)$ is the elementary amplitude for the scattering of a dipole of size $\vec{r}$ on the proton, $\vec{\Delta}$ denotes the transverse momentum lost by the outgoing proton (with $t=-\Delta^2$), $x$ is the Bjorken variable. The elementary elastic amplitude $\mathcal{A}_{q\bar q}$ can be related to the $S$-matrix element $S(x,r,b)$ for the scattering of a dipole of size $\vec{r}$ at impact parameter $\vec{b}$ \cite{MPS,KMW}:
\begin{eqnarray}
  \mathcal{A}_{q\bar q}(x,r,\Delta) = \mathrm{i}\,\int \dif^2\vec{b}\;\mathrm{e}^{-\mathrm{i}\vec{b}\cdot\vec{\Delta}}\,2\left[1-S(x,r,b)\right].
  \label{eq:smatrix}
\end{eqnarray}

As one has a real photon at the initial state, only the transversely polarized overlap function contributes to the cross section.  Summed over the quark helicities, for a given quark flavor $f$ it is given by \cite{MW},
\begin{eqnarray}
  (\Psi_{\gamma^*}^*\Psi_{\gamma})_{T}^f & = & \frac{N_c\,\alpha_{\mathrm{em}}e_f^2}{2\pi^2}\left\{\left[z^2+\bar{z}^2\right]\varepsilon_1 K_1(\varepsilon_1 r) \varepsilon_2 K_1(\varepsilon_2 r) \right.\nonumber \\
& + &  \left. m_f^2 K_0(\varepsilon_1 r) K_0(\varepsilon_2 r)\right\},
  \label{eq:overlap_dvcs}
\end{eqnarray}
where we have defined the quantities $\varepsilon_{1,2}^2 = z\bar{z}\,Q_{1,2}^2+m_f^2$ and $\bar{z}=(1-z)$. Accordingly, the photon virtualities are $Q_1^2=Q^2$ (incoming virtual photon) and $Q_2^2=0$ (outgoing real photon). In what follows we set the quark masses as $m_{u,d,s}=0.14$ GeV for the light quarks and $m_c=1.4$ GeV for the charm quark. It should be noticed that the quark mass regularizes the wave function of the real photon. Disregarding the real part of amplitude, the elastic diffractive cross section is then given by,
\begin{eqnarray}
  \frac{\dif\sigma^{\gamma^* p\rightarrow \gamma p}}{\dif t}
  & = & \frac{1}{16\pi}\left\lvert\mathcal{A}^{\gamma^* p\rightarrow \gamma p}(x,Q,\Delta)\right\rvert^2
  \label{eq:xvecm1}
\end{eqnarray}

\begin{figure}[t]
\includegraphics[scale=0.47]{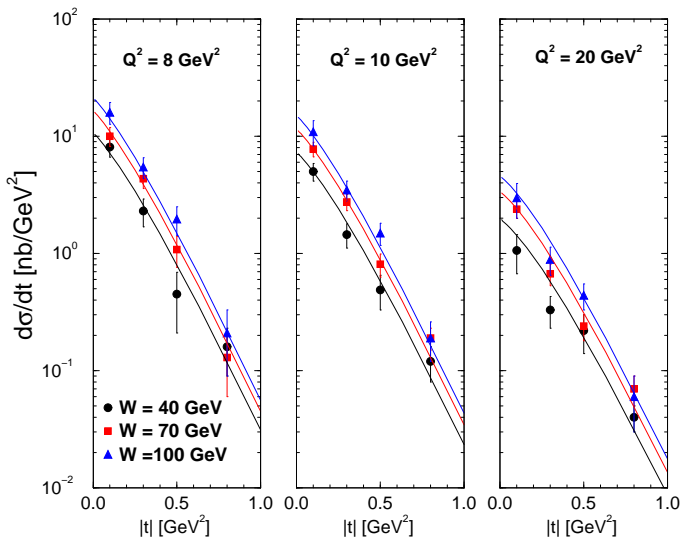}
\caption{The DVCS differential cross section, $d\sigma^{\gamma^*p\rightarrow \gamma p}/d|t|$, as a function of $|t|$ for distinct energies at moderate photon virtualities. MPS model compared to DESY-HERA data.}
\label{fig:2}
\end{figure}

For the DVCS on nucleons, we take into account saturation models which successfully describe exclusive processes at high energies. In particular, we consider the non-forward saturation model of Ref. \cite{MPS} (hereafter MPS model), which captures the main features of the dependence on energy,  virtual photon virtuality and momentum transfer $t$. An important result about the growth of the dipole amplitude towards the saturation regime is the geometric scaling regime \cite{gsincl,travwaves}. This geometric scaling property can be extended to the case
of non zero momentum transfer \cite{MAPESO}, provided $r\Delta\ll 1$. In the MPS model, the elementary elastic amplitude for dipole interaction is given by,
\begin{eqnarray}
\label{sigdipt}
\mathcal{A}_{q\bar q}(x,r,\Delta)= 2\pi R_p^2\,e^{-B|t|}N \left(rQ_{\mathrm{sat}}(x,|t|),x\right),
\end{eqnarray}
with the asymptotic behaviors $Q_{\mathrm{sat}}^2(x,\Delta)\sim
\max(Q_0^2,\Delta^2)\,\exp[-\lambda \ln(x)]$. Specifically, the $t$ dependence of the saturation scale is parametrised as
\begin{eqnarray}
\label{qsatt}
Q_{\mathrm{sat}}^2\,(x,|t|)=Q_0^2(1+c|t|)\:\left(\frac{1}{x}\right)^{\lambda}\,, \end{eqnarray}
in order to interpolate smoothly between the small and intermediate transfer
regions. The form factor $F(\Delta)=\exp(-B|t|)$ catches the transfer dependence of the proton vertex, which is factorised from the
projectile vertices and  does not spoil the geometric scaling properties. For the parameter $B$ we use the value $B=3.754$ GeV$^{-2}$ taken from Ref. \cite{MPS} (this parameter is reasonably stable in the phenomenological fits of MPS model). Finally, the scaling function $N$ is obtained from the forward saturation model
\cite{Iancu:2003ge}, whose functional form is given by,
\begin{equation}
\label{eq:bcgc}
 N(x,\,r) =\begin{cases}
  \mathcal{N}_0\left(\frac{rQ_{\mathrm{sat}}}{2}\right)^{2\left(\gamma_s+\frac{1}{\kappa\lambda Y}\ln\frac{2}{rQ_{\mathrm{sat}}}\right)} & :\quad rQ_{\mathrm{sat}}\le 2\\
  1-\mathrm{e}^{-A\ln^2(BrQ_{\mathrm{sat}})} & :\quad rQ_{\mathrm{sat}}>2
  \end{cases},
\end{equation}
where $Y=\ln(1/x)$ and the parameters for $N$ are taken from the original CGC model \cite{Iancu:2003ge}.
In order to study the sensitivity to a model dependence, we compare the MPS saturation model to impact parameter saturation model \cite{KMW}(hereafter b-SAT). In this saturation model, the elementary dipole-nucleon scattering amplitude is written in the impact parameter space as referred in Eq. (\ref{eq:smatrix}). In particular, in the b-SAT model the $S$-matrix element is given by \cite{KMW},
\begin{eqnarray}
S(x,r,b)=\exp\left[-\frac{\pi^2}{2N_c}r^2\alpha_S(\mu^2)\,xg(x,\mu^2)\,T(b)\right],\nonumber  \\
\label{bsatelem}
\end{eqnarray}
where the scale $\mu^2$ is related to the dipole size $r$ by $\mu^2=4/r^2+\mu_0^2$.  The gluon density, $xg(x,\mu^2)$, is evolved from a scale $\mu_0^2$ up to $\mu^2$ using LO DGLAP evolution without quarks. The initial gluon density at the scale $\mu_0^2$ is taken in the form $ xg(x,\mu_0^2) = A_g\,x^{-\lambda_g}\,(1-x)^{5.6}$. The proton shape function $T(b)$ is normalized so that $\int\!\dif^2\vec{b}\;T(b) = 1$ and one considers a Gaussian form for $T(b)$, that is, $ T(b) = \frac{1}{2\pi B_G}\mathrm{e}^{-\frac{b^2}{2B_G}}$ where $B_G=4$ GeV$^{-2}$ \cite{KMW}.

It should be noticed that in the DVCS process we need to use off-diagonal gluon distribution, since the exchanged gluons carry different fractions $x$ and $x^\prime$ of the proton's light-cone momentum. In order to take into account this skewedness correction, in the limit that $x^\prime \ll x \ll 1$, the elastic differential cross section should be multiplied by a factor $R_g^2$, given by \cite{Shuvaev:1999ce}
\begin{eqnarray}
\label{eq:Rg}
  R_g(\lambda_e) = \frac{2^{2\lambda_e+3}}{\sqrt{\pi}}\frac{\Gamma(\lambda_e+5/2)}{\Gamma(\lambda_e+4)}, \nonumber \\
  \quad\text{with} \quad \lambda_e \equiv \frac{\partial\ln\left[\mathcal{A}(x,\,Q^2,\,|t|)\right]}{\partial\ln(1/x)},
\end{eqnarray}
which gives an important contribution mostly at large virtualities. In our approximated expression for the skewedness correction, Eq. (\ref{eq:Rg}), we call attention that $\lambda_e$ is the effective power exponent of the imaginary part of amplitude. In addition, we will take into account the correction for real part of the amplitude, using dispersion relations $Re {\cal A}/Im {\cal A}=\mathrm{tan}\,(\pi \lambda_e/2)$. In the MPS model, the skewedness correction is absorbed in the model parameters and only real part of amplitude will be considered.

In order to cross check the results for DVCS on nucleons using the distinct saturation models referred  above, in Fig. \ref{fig:1} we present the differential DVCS cross section as a function of momentum transfer. The numerical results are compared to recent DESY-HERA \cite{H1dvcs} results at energy $W=82$ GeV for different photon virtualities ($Q^2=5,\,15.5,\,25$ GeV$^2$). Solid lines represent the results for MPS model and long-dashed curves for b-SAT model. We verify that they differ for $|t|>0.5$ GeV$^{2}$, with b-SAT model overestimating the data in such region. The MPS model gives a better data description in the kinematical range of presented experimental measurements. In what follows we will use the MPS model as the baseline model for numerical calculations. Thus, in Fig. \ref{fig:2} it is presented the DVCS differential cross section for distinct energies ($W=40,\,70,\,100$ GeV) and photon virtualities ($Q^2=8,\,10,\,20$ GeV$^2$) where the curves correspond to MPS model compared to DESY-HERA data \cite{H1dvcs}. The agreement is reasonably good in a wide range on energies at moderate virtualities and it would be suitable for extrapolation for nuclear target.

Let us now present  a preliminary study on the nuclear DVCS using the color dipole formalism at small-$x$. In the situation when the recoiled nucleus is not detected, measurements of DVCS observables with nuclear targets involves the coherent and incoherent contributions. The coherent scattering corresponds to the case in which the nuclear target remains intact and it dominates at small $t$. The incoherent scattering occurs when the initial nucleus of atomic number $A$ transforms into the system of $(A-1)$ spectator bound/free nucleons and one interacting nucleon and it dominates at large $t$. In what follows we address both contributions to nuclear DVCS cross section.

Here, we start by the coherent (elastic) nuclear DVCS contribution, $\gamma^* A \rightarrow \gamma A$, where the recoiled nucleus is intact. The implementation of nuclear effects in such a process is relatively   simple within the color dipole formalism in the low $x$ region. The usual procedure is to consider the Glauber-Gribov formalism for nuclear absorption. In phenomenological models in which geometric scaling is present (as in MPS saturation model) the extrapolation for a nucleus target is simplified. Therefore, here we will rely on the geometric scaling property \cite{gsincl,travwaves} of the saturation models: such a scaling  means that  the total $\gamma^* p$
cross section at large energies is not a function of the two
independent variables $x$ and $Q$, but is rather a function of the
single variable $\tau_p = Q^2/Q_{\mathrm{sat}}^2(x)$ as shown
in Ref. \cite{gsincl}. That is, $\sigma_{\gamma^*p}(x,Q^2)=\sigma_{\gamma^*p}(\tau_p)$. In Refs. \cite{travwaves} it was shown that the geometric scaling observed in experimental data can be understood theoretically in the context of non-linear QCD evolution with fixed and running coupling. Recently, the high energy $\ell^{\pm}p$,  $pA$ and $AA$ collisions have been related through geometric scaling \cite{Armesto_scal}. Within the color dipole picture and making use of a rescaling of the impact parameter of the $\gamma^*h$ cross section in terms of hadronic target radius $R_h$, the nuclear dependence of the $\gamma^*A$ cross section is absorbed in the $A$-dependence of the saturation scale via geometric scaling. The relation reads as $\sigma^{\gamma^*A}_{tot}(x,Q^2)  =  \kappa_A\,\sigma^{\gamma^*p}_{tot}\,(Q_{\mathrm{sat},p} \rightarrow Q_{\mathrm{sat},A})$, where $\kappa_A = (R_A/R_p)^2$. The nuclear saturation scale was assumed to rise with the quotient of the transverse parton densities to the power $\Delta \approx 1$ and $R_A$ is the nuclear radius, $Q_{\mathrm{sat},A}^2=(A/\kappa_A)^{\Delta}\,Q_{\mathrm{sat},p}^2$. This assumption successfully describes small-$x$ data for $ep$ and $eA$ scattering using $\Delta =1.26$ and a same scaling curve for the proton and nucleus \cite{Armesto_scal}.

Following the arguments referred above we will replace $R_p \rightarrow R_A$ in Eq. (\ref{sigdipt}) and also $Q_{\mathrm{sat},p}^2 (x,t=0)\rightarrow (AR_p^2/R_A^2)^{\Delta}\,Q_{\mathrm{sat},p}^2(x,t=0)$ in Eq. (\ref{qsatt}). In case of $\Delta=1$ the previous replacement becomes the usual assumption for the nuclear saturation scale, $Q_{\mathrm{sat},A}^2=A^{1/3}\,Q_{\mathrm{sat},p}^2$. For simplicity, we replace the form factor $F(t)=\exp (-B|t|)$ in Eq. (\ref{sigdipt}) by the corresponding nuclear form factor $F_A(t)= \exp (-\frac{R_A^2}{6}|t|)$. That is, the elementary elastic dipole-nucleus scattering amplitude now reads as,
\begin{eqnarray}
\label{sigdipnuc}
\mathcal{A}_{q\bar q}^{\mathrm{nuc}}(x,r,\Delta)= 2\pi R_A^2\,F_A(t)\,N \left(rQ_{\mathrm{sat},\,A};\,x\right).
\end{eqnarray}

For precise calculation, the nuclear form factor has to be replaced by the realistic hard sphere profile function. For the b-SAT saturation model, we just replace the proton shape function $T(b)$ in Eq. (\ref{bsatelem}) by the corresponding nuclear profile $T_A(b)$ (Wood-Saxon). It should be stressed that we are considering the limit of long coherence time, that is $l_c\gg R_A$. For small dipoles the scaling function is proportional to the nuclear saturation scale, $N(r\ll1/Q_{\mathrm{sat}})\propto r^2Q_{\mathrm{sat,\,A}}^2 \approx A^{1/3}\,r^2Q_{\mathrm{sat},p}^2$. Thus, in a rough approximation the coherent DVCS cross section will take the form $\frac{d\sigma_{\mathrm{coh}}}{dt}\approx A^2 F_A^2(t)\frac{d\sigma_N}{dt}$. This means that in the coherent case the differential cross section scales as $A^2$ and has a steep momentum transfer dependence due to the nuclear form factor squared,  $F_A^2(t)=\exp (-\frac{R_A^2}{3}|t|)$. The total coherent cross section (integrated over $|t|$) has an approximate $A$ dependence given by $\sigma_{\mathrm{coh}}\propto A^{4/3}\sigma_N$.

\begin{figure}[t]
\includegraphics[scale=0.47]{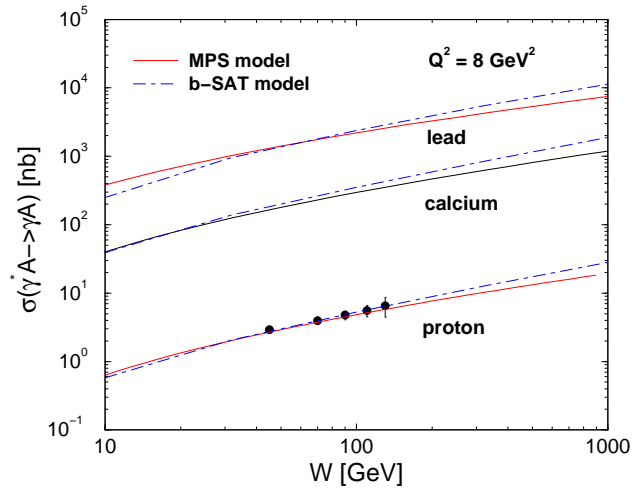}
\caption{The coherent DVCS integrated cross section as a function of photon-nucleus center of mass energy  at $Q^2=8$ GeV$^2$.}
\label{fig:3}
\end{figure}

Finally, we address the incoherent nuclear DVCS cross section. In diffractive incoherent (quasi-elastic) production of direct photons off nuclei, $\gamma^*A\rightarrow \gamma X$, one sums over all final states of the target nucleus except those which contain particle creation.  In order to compute the incoherent cross section we consider an approach involving the vector-dominance model (VDM) combined with the Glauber eikonal approximation. The Gribov corrections are calculated within the color dipole approach. Explicit  calculations can be found in Ref. \cite{KopNik} in the approximation of a short coherence (or production) length, $\ell_c$, when one can treat the creation of the colorless $q\bar{q}$ pair as instantaneous compared to the formation length, $\ell_f$, which is comparable with the nuclear radius $R_A$. In Ref. \cite{KopNik} the wave function formation is described by means of the light-cone Green function approach summing up all possible paths of the $q\bar{q}$ pair in the nucleus. The cross section for the incoherent production on nuclei can be derived under various conditions imposed by the coherent length.
We consider the high energy regime where the coherence length, $\ell_c=2\omega /Q^2$, is large such that $\ell_c \gg R_A$ ($\omega$ is the energy of the virtual photon in the rest frame of the nucleus). In this case, the transverse size of the $q\bar{q}$ wave packet is "frozen" by Lorentz time dilatation and the pairs do not fluctuate during propagation through the nucleus. The expression for the incoherent cross section is given by \cite{KopNik}:
\begin{eqnarray}
& & \left. \frac{d\sigma^{T,L}}{dt}\right|_{t=0}=\int d^2b\,T_A(b) \times \nonumber\\
& & \left|\left\langle \Psi_{\gamma}^{T,L}\left|  \sigma_{dip}(x,r)\exp\left[-\frac{1}{2}\sigma_{dip}(x,r)T_A(b)\right]\right| \Psi_{\gamma^*}^{T,L} \right\rangle\right|^2,\nonumber \\
\end{eqnarray}
where $T_A(b)=\int_{-\infty}^{+\infty}dz\,\rho(b,z)$ is the nuclear thickness function given by the integral of the nuclear density along the trajectory at a given impact parameter $b$. The quantity $\sigma_{dip}$ is the forward dipole-target elastic amplitude, that is $\sigma_{dip}(x,r)=\mathcal{A}_{q\bar q}\,(x,r,\Delta = 0)$. The light-cone wave functions for transverse and longitudinal photons at initial and final state are labeled by $|\Psi_{\gamma,\,\gamma^*}^{T,L}\rangle$.

For the total incoherent DVCS cross section we consider $\sigma_{\mathrm{incoh}}=|{\cal A}({\gamma^*A\rightarrow \gamma\,X})|^2/(16\pi B_{\mathrm{DVCS}})$, where $B_{\mathrm{DVCS}}$ is the $t$-slope in the DVCS on nucleons. The $t$-slope of the differential quasi-elastic cross section is the same as on a nucleon target. For our numerical estimations we use the experimental fit \cite{H1dvcs} to the slope, $B_{\mathrm{DVCS}}=A[1-B\log (Q^2/Q_0^2)]$, where $A=6.98\pm0.54$ GeV$^2$, $Q_0^2=2$ GeV$^2$ and $B=0.12\pm0.03$. The forward scattering amplitude is given by,
\begin{eqnarray}
\label{incoh}
& & \left|{\cal A}_{\gamma^*A\rightarrow \gamma\,X}(W,Q^2,t=0)\right|^2=\int d^2b\,T_A(b) \times \nonumber \\
& &\left|\int d^2r\int dz\, \Phi_{\gamma^*\gamma}\,\sigma_{dip}(x,r)\,\exp\left[ -\frac{1}{2}\sigma_{dip}(x,r)T_A(b)\right]  \right|^2.\nonumber\\
\end{eqnarray}

In the Eq. (\ref{incoh}), $\Phi_{\gamma^*\gamma}(z,r,Q^2;m_f)$ is the overlap function given by expression  in Eq. (\ref{eq:overlap_dvcs}). In a rough approximation, the incoherent DVCS cross section behaves like $\frac{d\sigma_{\mathrm{incoh}}}{dt}\approx A \frac{d\sigma}{dt}$, where $d\sigma/dt$ denotes the differential cross section for the quasi-free nucleon. The behavior on momentum transfer is slower in comparison to the coherent case and it is driven by the t-dependence of the cross section on quasi-free nucleons. In addition, it scales as $A$ in contrast to a $A^2$ scaling in the coherent case. In next section we compare the numerical calculations for the coherent and incoherent cross section for DVCS on nuclei.

\section{Phenomenology and numerical results}

In this section the numerical results for the DVCS cross section on nuclei are presented using the color dipole picture as referred in previous section.  The sensitivity to a different choice for the dipole cross section is also analyzed. Let us start by the coherent DVCS cross section. In Fig. \ref{fig:3} the integrated cross section is shown as a function of energy for fixed $Q^2=8$ GeV$^2$. The proton case is also presented for sake of comparison. We consider a light (calcium) and a heavy (lead) nucleus and distinct dipole cross sections (MPS and b-SAT models). The main difference between the models is the energy growth, which is simply understood from the different saturation scales and the QCD evolution in the b-SAT model. Thus, the b-SAT model (dot-dashed lines) produces a steep energy increasing in contrast to MSP model (solid lines). At high energies, the results for MPS model can be parameterized as $\sigma_{\mathrm{coh}}=\sigma(W_0)[W/W_0]^{\alpha}$ with the following parameters $\sigma(W_0)=293,\,(2151)$ nb, $\alpha=0.62,\,(0.55)$ for calcium (lead) and $W_0=100$ GeV. It is very clear the suppression for heavy nucleus in contrast to the nucleon case. The same parameterization produces the following results for the b-SAT model: $\sigma(W_0)=354,\,(2388)$ nb, $\alpha=0.73,\,(0.68)$ for calcium (lead).

\begin{figure}[t]
\includegraphics[scale=0.47]{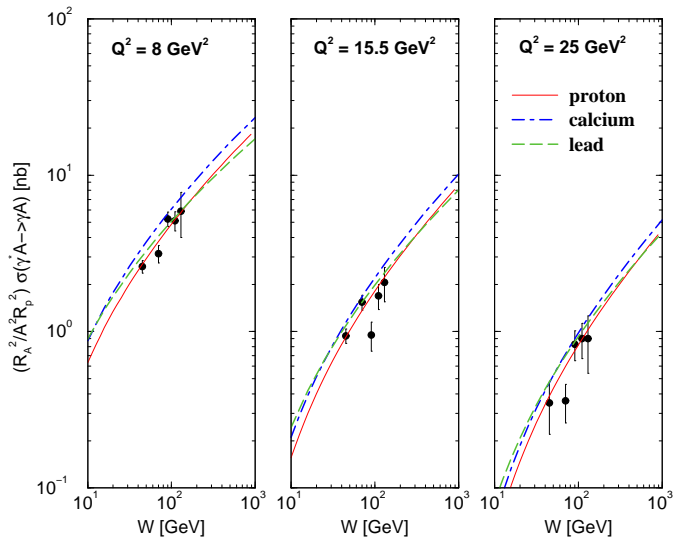}
\caption{The coherent DVCS cross section rescaled by factor $A^2R_p^2/R_A^2$. The DVCS cross section on nucleons is also presented.}
\label{fig:4}
\end{figure}

In order to study the $Q^2$-evolution, in Fig. \ref{fig:4} the cross section is shown as a function of energy for distinct virtualities $Q^2=8,\,15.5,\,20$ GeV$^2$. For sake of comparison, the DVCS cross section on nucleons (solid lines) is also presented and compared with experimental measurements from DESY-HERA. We consider calcium (dot-dashed lines) and lead (dashed lines) nuclei. The nuclear cross section has been rescaled by a factor $R_A^2/A^2R_p^2$ for illustration. The reason for that it is the  difficulty to compare the nuclear cross section to the Born term and the case $A=1$ does not match the DVCS cross section on proton (different $t$-slopes). For small size dipoles the amplitude, Eq. {\ref{sigdipnuc}), can be approximated by $\mathcal{A}_{q\bar q}\propto R_A^2\,Q_{\mathrm{sat,\,A}}^2r^2F_A(t)$. After overlapping, the scattering amplitude squared behaves like $|{\cal A}(s,t)|^2\propto R_A^4[Q_{sat\,A}^2/Q^2]^2\,F_A^2(t)$, which gives a integrated cross section $\sigma_{\mathrm{coh}}\propto R_A^2 Q_{\mathrm{sat\,A}}^4=R_A^2(AR_p^2/R_A^2)^{2\Delta}$. In the limit case where $\Delta = 1$,  $\sigma_{\mathrm{coh}}\propto A^2R_p^4/R_A^2$. It is verified a strong suppression for heavy nucleus at low $Q^2$. As the photon virtuality increases the corresponding suppression  diminishes. This fact is consistent with the general features of nuclear shadowing at small-$x$.
\begin{figure}[t]
\includegraphics[scale=0.47]{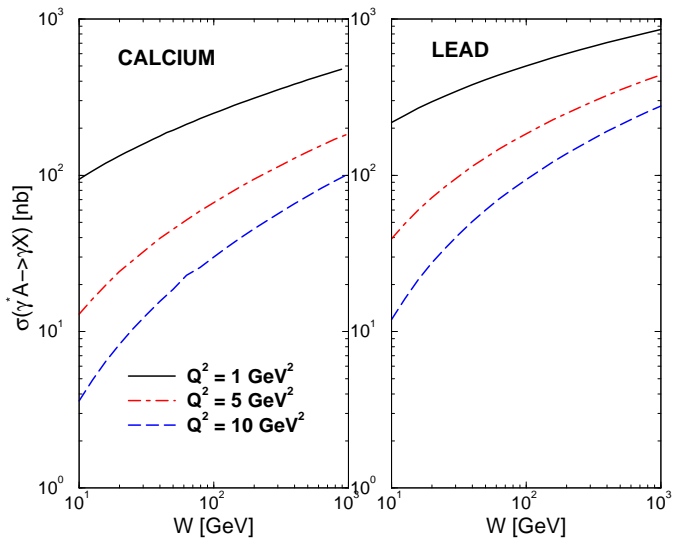}
\caption{The incoherent DVCS cross section for calcium and lead nucleus as a function of energy for fixed values of photon virtualities.}
\label{fig:5}
\end{figure}

Finally, in Fig. \ref{fig:5} the incoherent DVCS cross section is presented. The estimation is done for calcium (left panel) and lead (right panel) nucleus as a function of energy for representative values of photon virtualities ($Q^2=1,\,5,\,10$ GeV$^2$. The incoherent is suppressed by a factor 3-4 for calcium and by factor 5-6 for lead in comparison to the coherent case. This suppression can be understood from the different $A$-dependences of the integrated cross sections: the coherent DVCS cross section scales as $A^{4/3}$ whereas the incoherent cross section scales as $A$. Roughly speaking, the ratio incoherent/coherent scales as $A^{-1/3}$, which is consistent with the values found for the suppression. In order to illustrate the energy dependence of the incoherent DVCS cross section we parameterize it in the form $\sigma_{\mathrm{incoh}}=\sigma(W_0)[W/W_0]^{\alpha}$, with $W_0=100$ GeV.  For $Q^2=1$  GeV$^2$ one obtains $\sigma(W_0)=241,\,(480)$ nb, $\alpha=0.33,\,(0.27)$ for calcium (lead). It is verified that the effective energy exponent is a factor two smaller than the coherent case. This is directly associated to the strong exponential suppression for heavy nuclei appearing in Eq. (\ref{incoh}). That is, the dipole cross section attenuates with a constant absorption cross section. It should be stressed that in present calculations we assume the limit of large coherent length,  $\ell_c=2\omega /Q^2\gg R_A$, where approximation given by Eq. (\ref{incoh}) is valid.

\section{Summary}

Using the color dipole formalism, we studied the DVCS process on nucleons and nuclei. Such an approach is robust in describing a wide class of exclusive processes measured at DESY-HERA and at the experiment CLAS (Jeferson Lab.), like meson production, diffractive DIS and DVCS. The theoretical uncertainties are smaller in this case in contrast to the exclusive vector meson production as the overlap photon function are well determined. For the analysis of the nuclear case, we compare the geometric scaling based models to the usual Glauber-Gribov nuclear shadowing. We also provide estimations for the coherent and incoherent DVCS cross section, investigating their $A$-dependence. This is timely once DVCS off nuclei is a very promising tool for the investigation of the partonic structure of nuclei and it can be useful to clarify physics issues related to planned electron ion colliders (EIC's) as the LHeC.

\begin{acknowledgments}
 The author acknowledges the organizers of the {\it XXXVIII International Symposium on Multiparticle Dynamics -- ISMD2008} (Hamburg 15-20 September 2008), for their invitation and warm hospitality at Research Centre DESY,  where part of this work was performed. This work was supported by CNPq, Brazil.
\end{acknowledgments}

\end{document}